\documentclass{article}

\usepackage{PRIMEarxiv}

\usepackage[utf8]{inputenc} 
\usepackage[T1]{fontenc}    
\usepackage{hyperref}       
\usepackage{url}            
\usepackage{booktabs}       
\usepackage{amsfonts}       
\usepackage{nicefrac}       
\usepackage{microtype}      
\usepackage{lipsum}
\usepackage{fancyhdr}       
\usepackage{graphicx}       
\graphicspath{{media/}}     

\title{Wavelength and Phase Considerations for Multi-Pulse Plasma Generation of Terahertz
\thanks{\textit{\underline{Citation}}: 
\textbf{Moss, C. D., Sorenson, S.A., Johnson, J.A. DOI:000000/11111.}} 
}

\author{
  Clayton D. Moss (1), Shayne A. Sorenson (2), Jeremy A. Johnson (1,*) \\
 (1) Department of Chemistry and Biochemistry, Brigham Young University, Provo, UT 84602 \\
 (2) Department of Chemistry, Brigham Young University Idaho, Rexburg, ID, 83460\\
 *Corresponding author: jjohnson@chem.byu.edu}

\begin{document}
\maketitle

\begin{abstract}
We present a numerical study on plasma generation of THz radiation utilizing multiple light pulses of various wavelengths in an optical scheme that is readily achievable in a tabletop environment. To achieve coherent THz emission it is necessary to carefully consider all the wavelengths involved in a multi-pulse setup. Previous theoretical work has explored ideal waveforms and electric field symmetries for optimal efficiency in generating THz from plasma [Phys. Rev. Lett. {\bfseries 114} 183901 (2015)]. In practice such setups are quite delicate and prone to instability. We show that wavelength combinations with lower theoretical efficiency can more easily produce stable THz pulses in a tabletop environment combining readily available near-infrared wavelengths.
\end{abstract}

\section{Introduction}
Terahertz (THz) radiation is a useful spectroscopic tool, often employed to study ultrafast electronic and lattice dynamics. \cite{Hwang:15} One method of generating stable THz pulses for phase-resolved spectroscopy is focusing a two-color beam in air to generate a plasma. \cite{Cook:00,Andreeva:16,KimOE:07} One advantage of THz pulses generated from plasma is their extremely broad spectral bandwidths - a consequence of the highly nonlinear nature of the plasma generation process and the lack of any constraints imposed by a solid generation medium. Experimental and computational efforts have led to more optimal configurations, parameters of consideration include: beam focusing geometry\cite{Wang:11,Kuk:16}, fundamental pump wavelength\cite{Clerici:13,nikolaeva:22}, and using multiple colors of light beyond a standard two-color setup\cite{Martinez:15,Liu:20,alirezaee:20}. Notably, Martinez et al \cite{Martinez:15} proposed a "sawtooth" waveform made from multiple harmonics of light. While an ideal sawtooth shape is not feasible in a typical tabletop experiment, advances have been made to incorporate a third harmonic in plasma generation \cite{Liu:20,Ma:21}.

We and others have demonstrated that non-resonant three-color schemes can feasibly increase THz output in a tabletop setup with minimal additional equipment. \cite{Bagley,Sorenson,Vaicaitis:19,Ma:20} These involve adding an 800 nm beam to an existing two-color beam with an infrared (IR) fundamental. Often excess 800 nm light is available in a setup, e.g. discarded during down-conversion to IR or lost through reflective optics, even small amounts of 800 nm light can be used to increase plasma THz output. For example, we showed that when a 1450 nm fundamental is combined with its second harmonic (725 nm) adding additional 800 nm light can increase the THz output up to a factor of 30 for certain fluence combinations. \cite{Sorenson} 

We consider such experiments in this study, as diagrammed in Fig. 1a. In brief, an 800 nm laser is down-converted to the 1200-1800 nm range, which is more efficient for two-color plasma generation. \cite{Clerici:13,nikolaeva:22} The resulting IR beam is used as the fundamental wavelength and is focused through a second harmonic generation (SHG) crystal to form the two-color plasma. Often much of the initial 800 nm intensity persists after down-conversion and is discarded as "waste" light, as is the case in an optical parametric amplifier (OPA). The discarded waste light is recombined with the IR fundamental. In certain fluence and delay combinations the addition of the 800 nm beam increases THz output. An interesting result of these experiments is that inherent carrier-envelope phase (CEP) instabilities do not affect the coherence of the generated THz pulses. \cite{Sorenson,Ma:20} One possible explanation is that the frequency beating caused by using incommensurate wavelengths allows for this stability.

We expand on the experimental results of enhanced THz output from additional beams using the transverse photocurrent model to explore more possible generation schemes. In particular, we look at how THz output and stability are affected by the choice of fundamental IR wavelength. We discover that when the fundamental wavelength approaches twice 800 nm wavelength, the relative phase stability of all three pulses becomes vital. If the CEP is stable between the three beams, the highest theoretical gains in THz efficiency are realized. Interestingly, the relative CEP relationship between the three pulses becomes unimportant at certain wavelength combinations, showing that enhancement can still be achieved without relative CEP stability. These contrasting scenarios show that CEP stable laser systems are needed to obtain maximum THz generation enhancement, however specific wavelength combinations can be deployed to provide significant enhancement without a CEP-stable source. 

\begin{figure}[htbp]
\centering\includegraphics[width=7cm]{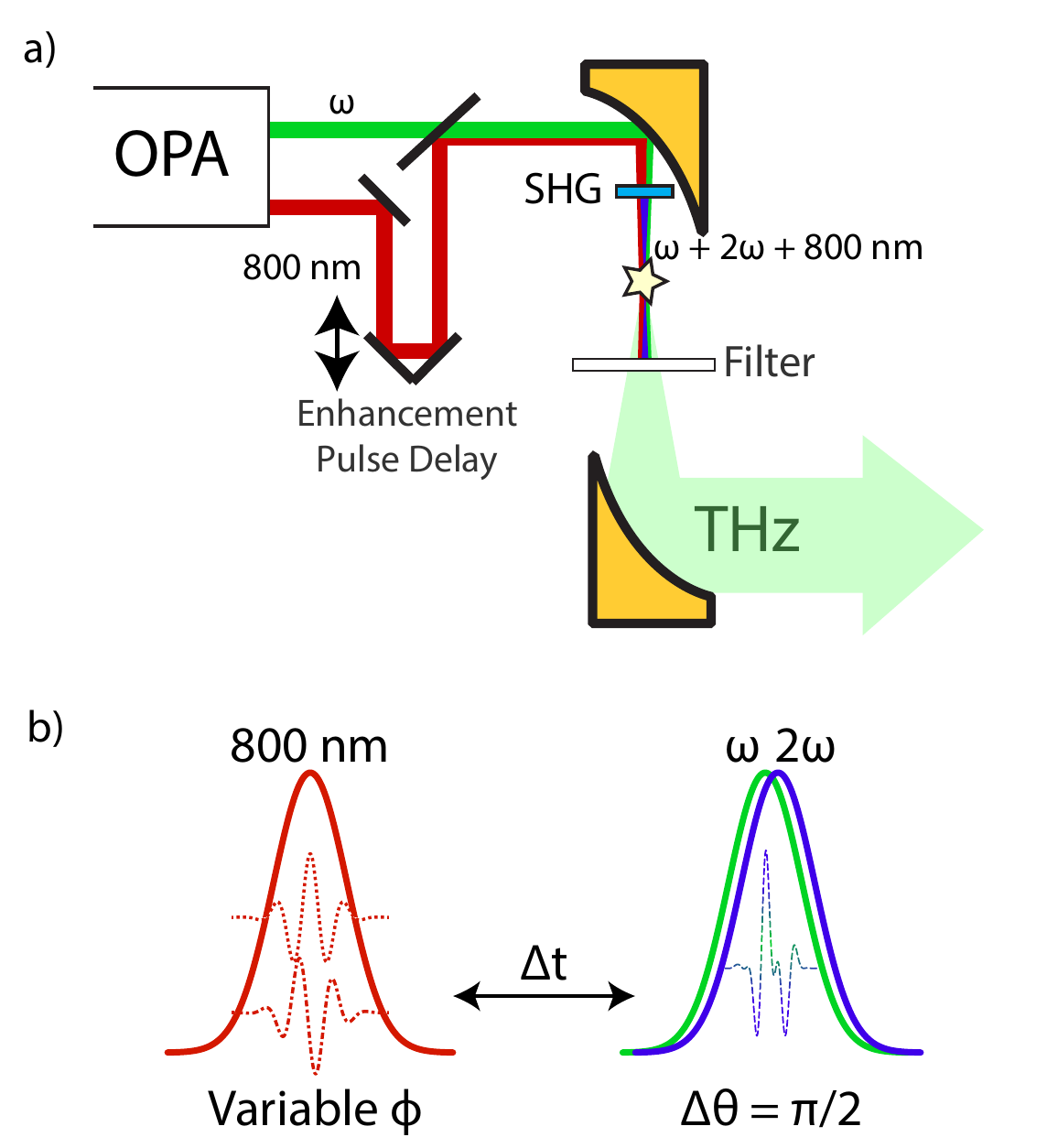}
\caption{a) Simulated experimental setup, adapted from \cite{Sorenson}. Separate beam lines are necessary to optimize the relative delay between the pulses before they are focused collinearly to make the plasma. b) Representation of simulation parameters, the primary variables are the IR fundamental frequency \(\omega\), relative delay \(\Delta t\), and 800 nm internal phase \(\phi\).} 
\end{figure}
\section{Methods}

We employ the photocurrent model of plasma-generated THz emission,\cite{KimPP:09} which has been shown to be the primary generation mechanism by more robust, computationally expensive particle-in-cell simulations. \cite{Andreeva:16} Modeling THz emission with only the photocurrent model has proven sufficient to provide qualitative analysis of experimental trends. \cite{Sorenson,Ma:21,Liu:20}
In the photocurrent model, the number of carriers in the plasma is calculated according to:
\begin{equation}
\frac{dN(t)}{dt} = w(t)[N_{g} -N(t)] - r_{t}N(t),
\end{equation}
with \(N(t)\) being the number of carriers at any given time \(t\), \(N_{g}\) being the initial number density of available carriers, and \(r_{t}\) representing the estimated recombination constant. Recombination occurs on longer time scales than generation, thus the recombination term can often be ignored. The tunneling rate \(w(t)\) is calculated in response to the magnitude of the driving field \(E(t)\) by:
\begin{equation}
    w(t) = \frac{\alpha}{\hat{E}(t)} exp(-\frac{\beta}{\hat{E}(t)}).
\end{equation}
Where \(\alpha = 4 \omega_a r_H^{5/2} \), \(\beta = (2/3) r_H^{3/2} \), \( \omega_a = 4.134 \times 10^{16} s^{-1}\) is the atomic frequency unit and \( r_H = U_{N_2}/U_H\) is the ionization potential of nitrogen gas (15.6 eV) relative to that atomic hydrogen (13.6 eV). We also define \( \hat{E}(t) = |E(t)|/E_a \) as the absolute value of the electric field of the laser in atomic units (\(E_a = 5.14 \times 10^{11}\) V/m). 
The emitted THz field is proportional to the derivative of the electron shift current that occurs in the plasma, which is simply the product of the carrier generation function and the electric field:
\begin{equation}
    E_{emit} \propto \frac{e^2}{m_{e}}E(t)N(t).
\end{equation}
From this modeled emission we calculate THz yield as the square root of the absolute value of the Fourier transform squared, with the frequency bounded from 0.1 to 10 THz. 

The experimental setup which we simulate in our calculations is depicted in Fig. 1. As performed in Sorenson et al \cite{Sorenson}, a variable IR fundamental is chosen and focused through an optimally aligned SHG crystal. The excess 800 nm light taken from the OPA is reintroduced along the same focusing line with variable relative delay. At certain delays and relative powers, the optimized THz generation of the IR beam is enhanced further by the addition of the 800 nm beam. A detailed look at the phase considerations of the driving pulse is shown in Fig. 1b. The infrared fundamental frequency (\(\omega\)) and its second harmonic (\(2\omega\)) have their relative delay fixed at \(\pi\)/2, this optimizes THz output and is achieved in practice by alignment of the SHG crystal.  The reference THz yield is calculated using only these two colors. The fundamental is varied over a range of 1100 nm to 2000 nm, representative of feasible driving frequencies from an OPA. This is then compared to output with the third 800 nm beam present. The relative delay is varied in addition to the internal phase of the pulse (\(\phi\)). For this study we keep the fluence of the IR fundamental constant, corresponding to a pulse energy of 240 \( \mu \)J, the fluence of the 800 nm beam corresponds to a pulse energy of 1930 \( \mu \)J (see SI of \cite{Sorenson}). All pulses are modeled using Gaussian beam profiles with 75 fs set as the full-width-half-max. 
The relative phase of the 800 nm pulse is of particular interest to this work, in an experimental setup where beams follow different optical paths it is difficult to ensure fixed relative phase due to optical jitter and fluctuations. Regardless of whether fixed relative phases between all three pulses can be achieved, our calculations consider both fixed and random phase scenarios. In the fixed case all pulses are CEP stable. To model the random phase scenario, we model the THz emission for a range of relative phase delays (\(\phi\)) and then average the current produced by each phase (See Fig. 2). Averaging the current across different phases replicates the experimental condition of shot averaging. 

\section{Results and Discussion}

The modeled results of both conditions with and without CEP stability are shown in Fig. 2 for a 1450 nm fundamental and a 1600 nm fundamental. As shown in Fig. 2a, at optimal relative delays the emitted THz pulse is enhanced on average by a phase-unstable 800 nm pulse due to more favorable carrier ionization and subsequent current generation. If the 800 nm beam arrives before the two-color beam (negative relative delays) THz emission is suppressed as carriers are liberated without the asymmetric push needed to form a net current drift. For a primary IR wavelength that is far away from being harmonic with the third pulse (1450 nm, blue dotted lines, see Fig. 2b) the specific phase of the 800 nm pulse doesn't matter at optimal delay and there is consistent enhancement when averaging over every possible phase. However, when the primary wavelength is commensurate (1600 nm, red dashed lines, see Fig. 2c) the outcomes vary greatly for different fixed phases of the 800 nm pulse ($\phi$). The overall symmetry of the driving pulse is much more sensitive to the addition of the third beam, and we see that the average result is less desirable for the commensurate (1600 nm) compared to the the non-commensurate (1450 nm) case. However, if the phase of the third beam can be held stable and suitably optimized, then the greatest potential gains in THz emission are possible. For example, the solid green line corresponding to a phase of 4\(\pi\)/10 in Fig. 2c exceeds a maximum enhancement factor of four, larger than the predicted outcome for 1450 nm (Fig 2b). 

\begin{figure}[htbp]
\centering\includegraphics[width=13cm]{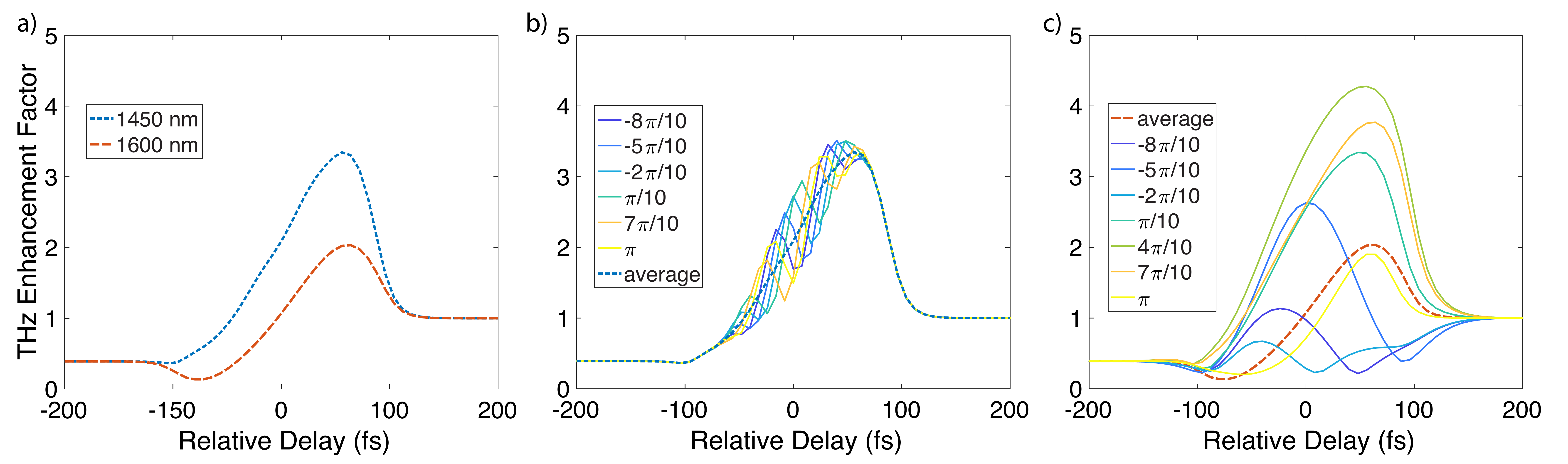}
\caption{We calculate THz enhancement factor as the ratio of the THz yields with the 800 nm pulse present and absent from the driving ionization field. Positive relative delays correspond with the 800 nm pulse arriving after the pulse containing the IR fundamental its second harmonic. a) Averaged, phase-unstable enhancement for the commensurate (1600 nm, red dashed line) and incommensurate (1450 nm, blue dotted line) cases. b) Individual fixed CEP stable (solid lines) and random average phase (dotted line) calculations for 1450 nm fundamental. There is little variation at optimal relative delay. c) Fixed CEP stable and random average phase (dashed line) calculations for 1600 nm fundamental. Each individual phase has drastically different outcomes.} 
\end{figure}

Note that the average in Fig. 2c (red dashed line) is different than the expected mean of the phases shown; this is because enhancement factor is calculated from scalar yields. Averaging accounts for both phase and magnitude derived from the electron current. In Fig. 3 we plot several calculated electron currents for a 1600 nm fundamental. While we use the derivative of the electron current to calculate emission (Eq. 3), the electron current, (in particular the trailing current tail seen after 150 fs in Fig. 3), gives a good sense of the magnitude and phase of a resulting THz pulse. The phase averaged current shift (blue dotted line) is not much larger than the calculated shift of the two-beam reference (black line). The optimal single fixed phase (light orange line) results in a much larger current and overall displacement, but different phases can create currents with displacement in the opposite direction (purple dashed line).  

\begin{figure}[htbp]
\centering\includegraphics[width=7cm]{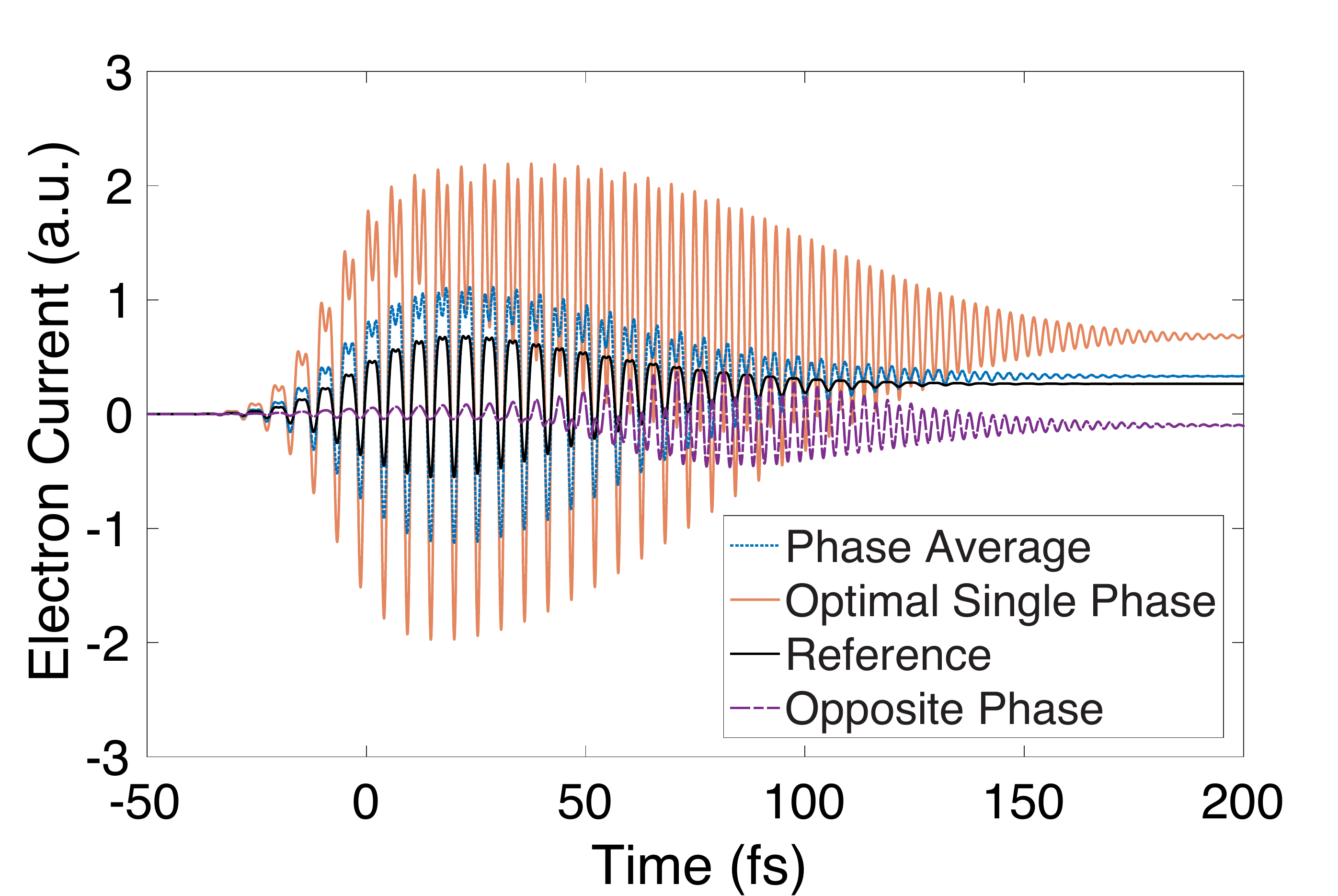}
\caption{Calculated electron photocurrents for a 1600 nm fundamental, which determine the phase and magnitude of resulting THz pulses. The low-frequency displacement from zero, observed clearly after 150 fs, is largely responsible for THz emission. The average of all possible 800 nm third beam phases (blue dotted line) is a small improvement over it not being present at all (black solid line). The optimal fixed-phase result (light orange line) is a significant improvement over the average. Certain phases can result in current shifts in the opposite direction (purple line), which explains the discrepancy between the optimal single phase and the phase-averaged cases.}
\end{figure}

In Fig. 4 we show the results from extending our model over a large range of IR fundamental wavelengths. Here, we report the enhancement factor for only the optimal relative time delay between driving pulses. This again assumes a fundamental with second harmonic and additional 800 nm. The largest gains in THz field strength are achieved at commensurate wavelengths, evidenced by the large feature centered around 1600 nm, but only in fixed-phase conditions. The individual fixed-phase results (thin lines) caution that reduced THz yield is equally possible, in the same situations that a fixed-phase scheme would have the highest gain. As the wavelength combinations become less commensurate, the optimal single-phase and phase-averaged cases converge. As seen at 1100 nm, 1400 nm, and 1800 nm, the difference between the CEP stable and random averaged cases is minimal. Conversely, we see larger discrepancies at full and fractional harmonics. \cite{Vvedenskii:14} We also note that while THz emission is more efficient at longer wavelengths it becomes increasingly difficult to create the optimal plasma sparks required for THz generation. \cite{Clerici:13} Our method accounts for increased Keldysh ionization tunneling due to the enhancement 800 nm pulse being present, but does not consider focusing difficulties that can arise using longer wavelengths.

Our results suggest two paths towards optimal THz generation beyond the two-color scheme. To achieve the highest theoretical THz yields it is necessary to have commensurate colors with carefully chosen relative phases for all beams. This allows for the pursuit of previously discussed optimal driving field shapes, such as the sawtooth waveform. \cite{Martinez:15} It is worth noting that in cases with fewer beams different phase combinations may be more favorable. \cite{alirezaee:20} However, if phase stability cannot be achieved experimentally, choosing incommensurate colors can still boost THz output while avoiding CEP stability concerns.

\begin{figure}[htbp]
\centering\includegraphics[width=7cm]{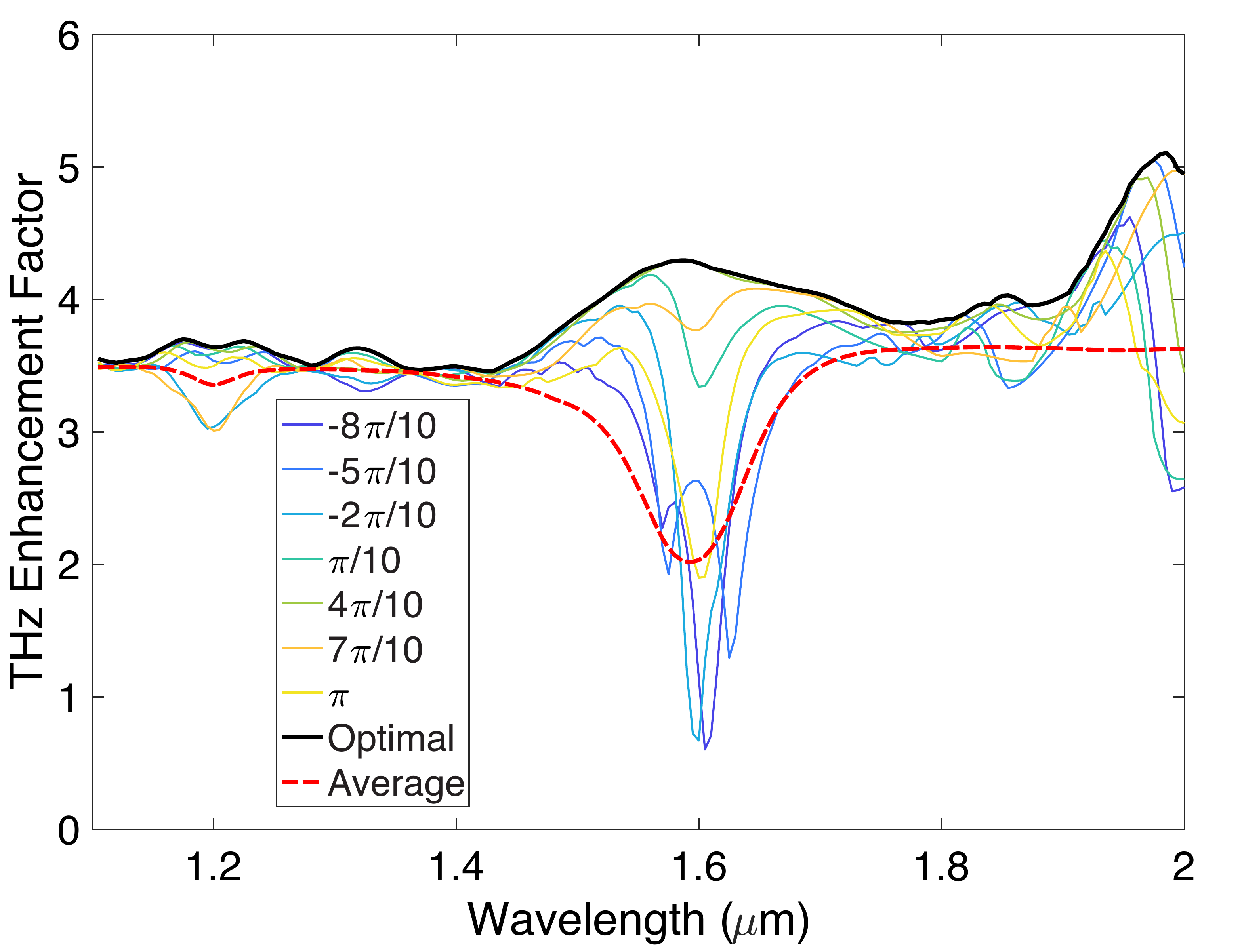}
\caption{THz enhancement at optimal phase delay plotted against fundamental IR frequency for a phase averaged (dashed red line) case and selected individual phases. The heavy black line shows the best result possible with a set, single-phase. The largest discrepancy between the single, stable pulse and averaged, unstable phase scenarios occurs at commensurate wavelengths. Smaller features correspond with fractional harmonics.}
\end{figure}

\section{Conclusions}

In conclusion we hope to bridge optimal theoretical generation scenarios and practical experimental considerations in multi-beam plasma generation of THz. We demonstrate that non-optimized, incommensurate wavelength scenarios can give better results within practical constraints that make relative CEP stability between driving pulses impossible. As multi-harmonic THz plasma generation schemes continue to improve we hope to encourage flexibility and creativity to achieve stable THz output. 

\bibliographystyle{unsrt}  
\bibliography{3colortry1}

\end{document}